\documentclass[12pt]{article}
\newlength{\vshift}
\usepackage{graphicx}
\newlength{\hshift}

\usepackage{amssymb}
\usepackage{amsmath}
\usepackage{amsthm}
\usepackage{amsopn}
\def\beq{\begin{equation}}
\def\eeq{\end{equation}}
\def\bea{\begin{eqnarray}}
\def\eea{\end{eqnarray}}

\title{On the Newtonian Anisotropic Configurations}
\author{F. Shojai$^{1,2}$, M.R. Fazel$^1$, A. Estepanian$^1$, M. Kohandel$^3$\\ $^1$Department of Physics, University of Tehran,\\ Tehran, Iran.\\$^2$Institute for Studies in Theoretical Physics and Mathematics (IPM),\\ Tehran, Iran.\\ $^3$ ‫Department ‪of‬‬ ‫‪Sciences,‬‬ ‫‪Alzahra‬‬ ‫‪University,\\‬‬ ‫‪Tehran, Iran‬‬.}
\date{}
\begin{document}
\maketitle
\abstract{In this paper we are concerned with the effects of anisotropic pressure on the boundary conditions of anisotropic Lane-Emden equation and homology theorem. Some new exact solutions of this equation are derived. Then some of the theorems governing the Newtonian perfect fluid star are extended taking the anisotropic pressure into account.}
\section{Introduction}
The Newtonian theory of stellar structure is usually used to describe the stars with not extremely high densities \cite{her2013,dev,chandra,horedt}. Even for high density compact objects like white dwarfs and neutron stars, applying Newtonian gravity leads to acceptable results comparable to the results of the relativistic models \cite{rod}.

The contents of the star is often modelled by a perfect fluid with an equation of state relating the mass density $\rho$ and pressure $P$. Among all possible choices of this equation, the polytrope equation of state, $P=K\rho^{\gamma}$ is of great importance for many astrophysical situations \cite{her2013, chandra, horedt} where $K$ and $\gamma=1+\frac{1}{n}$ are both constants called polytropic constant and polytropic exponent, respectively. ($n$ is called the polytropic index).
The polytrope equation is the simplest equation of state useful for a wide range of fluid densities. Moreover it leads to the scale symmetry and therefore introducing  homology invariants \cite{horedt}.

On the other hand there are some phenomena leading to the local anisotropy of pressure inside a self gravitating system. For low density objects, the Newtonian approximation is valid. Some of the physical mechanisms for anisotropy in this regime are \cite{her1997}:
\begin{itemize}
  \item When there is an anisotropic velocity distribution in a collisionless gas \cite{velocity}, the radial pressure satisfies Jeans equation:
\begin{equation}
\frac{dP_r}{dr}=\rho \frac{d\phi}{dr}+\frac{2}{r}\Delta(r)
\label{A1}\end{equation}
where $\phi$ is the Newtonian gravitational potential of the fluid and we have considered a static and spherically symmetric distribution of matter. $\Delta=P_t-P_r$, where $P_r$ and $P_t$ are the radial and tangential pressure respectively. In this case $\Delta$ measures the anisotropy of velocity distributions:
\begin{equation}
\Delta=\rho(\langle V^2_r\rangle -\langle V^2_\phi \rangle ) 
\end{equation} 
where $\langle V^2_r\rangle$ and $\langle V^2_\phi \rangle$ are radial and azimuthal velocity dispersions. (Note that according to the spherical symmetry assumption we have $\langle V^2_\phi \rangle=\langle V^2_\theta\rangle$). Equation (\ref{A1}) describes the hydrostatic equilibrium of a self gravitating object. 

To produce anisotropic velocity distribution, consider a galactic halo of fermionic dark matter. The conservation of angular momentum of neutrinos streaming into the halo, leads to anisotropic pressure \cite{fer}.
\item For a slowly rotating system \cite{horedt,her1997}, the equation of hydrostatic equilibrium in the first order is given by equation (\ref{A1}) in which $\Delta=-\frac{1}{3}\rho\omega^2r^2$ where 
$\omega$ is the angular velocity.
\item For a mixture of two non interacting perfect fluids, the energy-momentum tensor is that of an anisotropic fluid in which $\rho$, $P_r$ and $P_t$ are some functions of the mass densities and pressure of each component \cite{non}.
\item For a low mass charged white dwarf, the repulsive electrostatic force can be regarded as a source of anisotropy \cite{low}.
\end{itemize}
According to the above cases, even in the Newtonian regime, generally we are dealing with two components of pressure. 

Here we shall focus on the Newtonian anisotropic polytropes. In the next section, first we derive the anisotropic version of dimensionless Newtonian hydrostatic equation, Lane-Emden equation  and it's boundary conditions. We extend the homology theorem for anisotropic polytropes and then in section 3 we derive a new set of analytical solutions. There are some exact solutions of the Lane-Emden equation for special values of barotropic index in arbitrary spatial dimension \cite{horedt}. In this paper, we assume that the form of the Lane-Emden equation, doesn't change after introducing the anisotropy factor. Thus we find the anisotropic modified form of the existing isotropic solutions by this ansatz.  The authors of \cite{cos}, propose another ansatz on $\Delta$ in the modelling of relativistic stars. This is used  in \cite {her2013} to generate some anisotropic Newtonian polytrope solutions from isotropic solutions. Other various assumptions on $\Delta$ can be found in \cite{dev}. In section 4, we focus on some integral theorems on the physical quantities of a star in the hydrostatic equilibrium obtained by Chandrasekhar in his book \cite{chandra}. These are derived from Newtonian equations directly without any special assumption on the internal structure of the star or it's equation of state. These theorems contain some inequalities on the central pressure, the gravitational potential and the mean gravitational acceleration of a star. Here we extend these theorems to the general anisotropic equilibrium configuration. 
\section{Anisotropic polytropes}
The hydrostatic equilibrium of a star is governed by equation (\ref{A1}). The Newtonian potential is related to the density of the fluid by Poisson's equation:
\begin{equation}
\nabla^2 \phi =-4\pi G\rho.
\label{A2}\end{equation}
Considering the fluid obeys the polytropic equation of state along with the equation
(\ref{A1}), one gets:
\begin{equation}
\phi(r)-\phi_0=K(n+1)\left(\rho^{\frac{1}{n}}-\rho_0^{\frac{1}{n}}\right)-\int_0^r \frac{2\Delta(x)}{x\rho(x)}dx
\label{A4}\end{equation}
where $\rho_0$ and $\phi_0$ denote the central density and gravitational 
potential. We take $\phi_0=0$. Substituting equation (\ref{A4}) into equation (\ref{A2}) leads to the following expression which is the fundamental equation of equilibrium:
\begin{equation}
K(n+1)\nabla^2 \rho^{\frac{1}{n}}-\frac{1}{r^{N-1}}\frac{d}{dr}\left(r^{N-1}\frac{2\Delta(r)}{r\rho(r)}\right)=-4\pi G\rho
\label{A5}
\end{equation}
where $N$ is the dimension of space. 

Introducing the new dimensionless variable
$\xi=\left(\pm \frac{(n+1)K}{4\pi G {\rho_0}^{1-\frac{1}{n}}}\right)^{-\frac{1}{2}}r$, and with the following definitions
\begin{equation}
\rho ={\rho}_0 \theta^n, \hspace{1cm} P_r=P_0 {\theta}^{n+1}
\label{A6}\end{equation}
equation (\ref{A5}) takes the following form
\begin{equation}
\theta''+\frac{N-1}{\xi}\theta'-\frac{2}{P_0 (n+1)\xi\theta^n}\left[\Delta'+\frac{N-2}{\xi}\Delta-n\frac{\theta'}{\theta}\Delta\right]=\pm\theta^n.
\label{A7}\end{equation}
where prime denotes differentiation with respect to the new radial coordinate, $\xi$, and the plus and minus signs correspond to $-\infty<n<-1$, $-1<n<+\infty$  respectively. Note that $\theta=\theta(r)$ is a dimensionless function. Equation (\ref{A7}) is the anisotropic version of the well-known Lane-Emden equation. 

For $n=-1$, the polytrope equation of state gives $P_r=K$ if $\rho\neq 0$. Thus according to equation (\ref{A4}), the term involving $\Delta$ only contributes to the gravitational potential and the Poisson's equation gives:
\begin{equation}
\Delta=\frac{2\pi G\rho}{r^{N-2}}\int_0^r x^{N-1} \rho(x)dx
\label{AD}
\end{equation}
This shows that the case with $n=-1$ is a special case which must be treated separately.  Whereas for the isotropic case $(\Delta=0)$, one gets $\rho=0$ and therefore, $n=-1$ is excluded from the further study for isotropic fluids.

The other important case arises when $n=\pm \infty$ and therefore the polytrope equation of state reduces to
\begin{equation}
P_r=K\rho.
\label{AA}\end{equation}
Replacing this into (\ref{A1}), we obtain
\begin{equation}
\phi-\phi_0=K\ln\left(\frac{\rho}{\rho_0}\right)-\int^r_0\frac{2\Delta(x)}{\rho(x) x}dx
\label{AB}\end{equation}
The combination of the above expression when $\phi_0=0$ and (\ref{A2}) gives
\begin{equation}
\theta''+\frac{N-1}{\xi}\theta'+\frac{2}{P_0\xi}e^{\theta}\left(\Delta'+\frac{N-2}{\xi}\Delta+\Delta \theta'\right)=e^{-\theta}
\label{AC}\end{equation}
where $\rho=\rho_0 e^{-\theta}$, $P_r=P_0 e^{-\theta}$ and $\xi=\left(\frac{K}{4\pi G\rho_0}\right)^{-\frac{1}{2}}r$.

Either the Lane-Emden equation (\ref{A7}) (for $n\neq -1,\pm \infty$) or (\ref{AC}) (for $n=\pm \infty$), contains two unknown functions, $\theta(\xi)$ and $\Delta(\xi)$. Thus we need another equation to obtain a closed system of equations.

Before describing our procedure to introduce the other equation, let us assume that $\Delta$ is a given function. To obtain a unique solution for $\theta$ one has to specify two boundary conditions. The first is $\theta(0)=1$ (for $n\neq -1, \pm \infty$) and $\theta(0)=0$ (for $n=\pm \infty$). In order to obtain the other condition, let's integrate (\ref{A2}):
\begin{equation}
\frac{d\Phi}{dr}=-r^{1-N}{\int_0}^r 4\pi G\rho x^{N-1}dx
\label{A8}\end{equation}
Combination of the above relation with (\ref{A1}) and (\ref{A6}) , produces
\begin{equation}
\frac{d\theta}{d\xi}|_{\xi\to 0}=-\frac{\xi}{N}|_{\xi\to 0}+\frac{2}{P_0(n+1)}\lim_{\xi \to 0}\left(\frac{\Delta(\xi)}{\xi}\right).
\label{A9}\end{equation}
A similar calculation for $n=\pm \infty$ leads to:
\begin{equation}
\frac{d\theta}{d\xi}|_{\xi\to 0}=\frac{\xi}{N}|_{\xi\to 0}-\frac{2}{P_0}\lim_{\xi \to 0}\frac{\Delta(\xi)}{\xi}
\label{AE}\end{equation}
Relations (\ref{A9}) and (\ref{AE}) lead to the extensions of the Chandrasekhar's theorem \cite{chandra} for anisotropic polytrope:

\underline{\textbf{Theorem I:}}

The finite solutions of anisotropic Lane-Emden equation at the origin have to satisfy $\frac{d\theta}{d\xi}|_{\xi=0}=\frac{2}{P_0(n+1)}\lim_{\xi \to 0}\left(\frac{\Delta(\xi)}{\xi}\right)$ if $n\neq -1, \pm \infty$ and $\frac{d\theta}{d\xi}|_{\xi=0}=\frac{2}{P_0}\lim_{\xi \to 0}\frac{\Delta(\xi)}{\xi}$ if $n=\pm \infty$.

Another important theorem is a generalization of the homology theorem \cite{horedt}:

\underline{\textbf{Theorem II:}}

If the anisotropic Lane-Emden equation is satisfied by $\theta(\xi)$ and $\Delta(\xi)$ then $A^{\frac{2}{n-1}} \theta(A\xi)$ and $A^{\frac{2(n+1)}{1-n}}\Delta(A\xi)$ satisfy the equation (\ref{A7}), and $\theta(A\xi)-\ln A^2$ and $\frac{1}{A^2}\Delta(A\xi)$ satisfy the equation (\ref{AC}), where $A$ is a constant. 

The proof is straightforward. It can be easily obtained by substituting these expressions in (\ref{A7}) and (\ref{AC}). Thus we obtain a set of solutions parametrized by $A$.

One usual approach to solve (\ref{A7}) (or (\ref{AC})) with boundary conditions obtained in theorem I, is to suppose a special form for $\Delta$. Then putting it into (\ref{A7}) (or (\ref{AC})) yields a differential equation for $\theta$. If this procedure leads to smooth and non-negative pressures and density for the fluid, then they can be regarded as the physical solutions. However, knowing the function $\Delta$,  equation (\ref{A7}) (or (\ref{AC})) is a non-linear differential equation which in general one does not expect to obtain an analytical solution. Here we introduce a heuristic approach to get some analytical solutions. In order to do this, we assume that the anisotropy factor has no influence on the form of 
isotropic Lane-Emden equation. This means that $\Delta$ modifies only the coefficients of isotropic Lane-Emden equation. Hereafter, we adopt the minus sign on the right-hand side of Lane-Emden equation (\ref{A7}) (or (\ref{AC})) since the plus sign leads to an imaginary radial coordinate \cite{horedt}. In the following we shall consider two cases:

\underline{\textbf{Case I:}}

$\Delta$ modifies the coefficient of $\theta'$ in the isotropic Lane-Emden equation . In this case equation (\ref{A7}) can be considered as two separated equations as below
\begin{eqnarray}
\frac{-2}{P_0(n+1)\theta^n}\left[\Delta'+\frac{N-2}{\xi}\Delta-n\frac{\theta'}{\theta}\Delta\right]=N_1\theta'  \label{A10}\\
\theta''+\frac{N_1+N-1}{\xi}\theta'=-\theta^n \label{A11}\hspace{1.5cm}
\end{eqnarray}
where $N_1$ is an arbitrary constant. Equation (\ref{A10}) is a first order differential equation with the solution
\begin{equation}
\Delta(\xi)=-\frac{(n+1)P_0N_1\theta^n}{2\xi^{N-2}}\int \theta'\xi^{N-2}d\xi
\label{A12}\end{equation}

Equations (\ref{A11}) and (\ref{A12}) with the boundary conditions (\ref{A9}) and $\theta(0)=1$, give the solutions of anisotropic polytropes. 

\underline{\textbf{Case II:}}

$\Delta$ modifies the coefficient of $\theta^n$ in the right hand side of the isotropic Lane-Emden equation.
\begin{eqnarray}
\frac{-2}{P_0(n+1)\theta^n\xi}\left[\Delta'+\frac{N-2}{\xi}\Delta-n\frac{\theta'}{\theta}\Delta\right]=N_2\theta^{n} \label{AF}\\
\frac{d^2\theta}{d\eta^2}+\frac{N-1}{\eta}\frac{d\theta}{d\eta}=-\theta^n \hspace{1.1cm} \label{AG}
\end{eqnarray}
where $\eta=\sqrt{(N_2+1)}\,\ \xi$, $N_2>-1$. The solution of equation (\ref{AF}) is
\begin{equation}
\Delta(\xi)=\frac{-(n+1)P_0N_2\theta^n}{2(N_2+1)\xi^{N-2}}\int \theta^n \xi^{N-1}d\xi
\label{AH}\end{equation}

\begin{figure}[tbp]
\centering
\fbox{\includegraphics[scale=0.5]{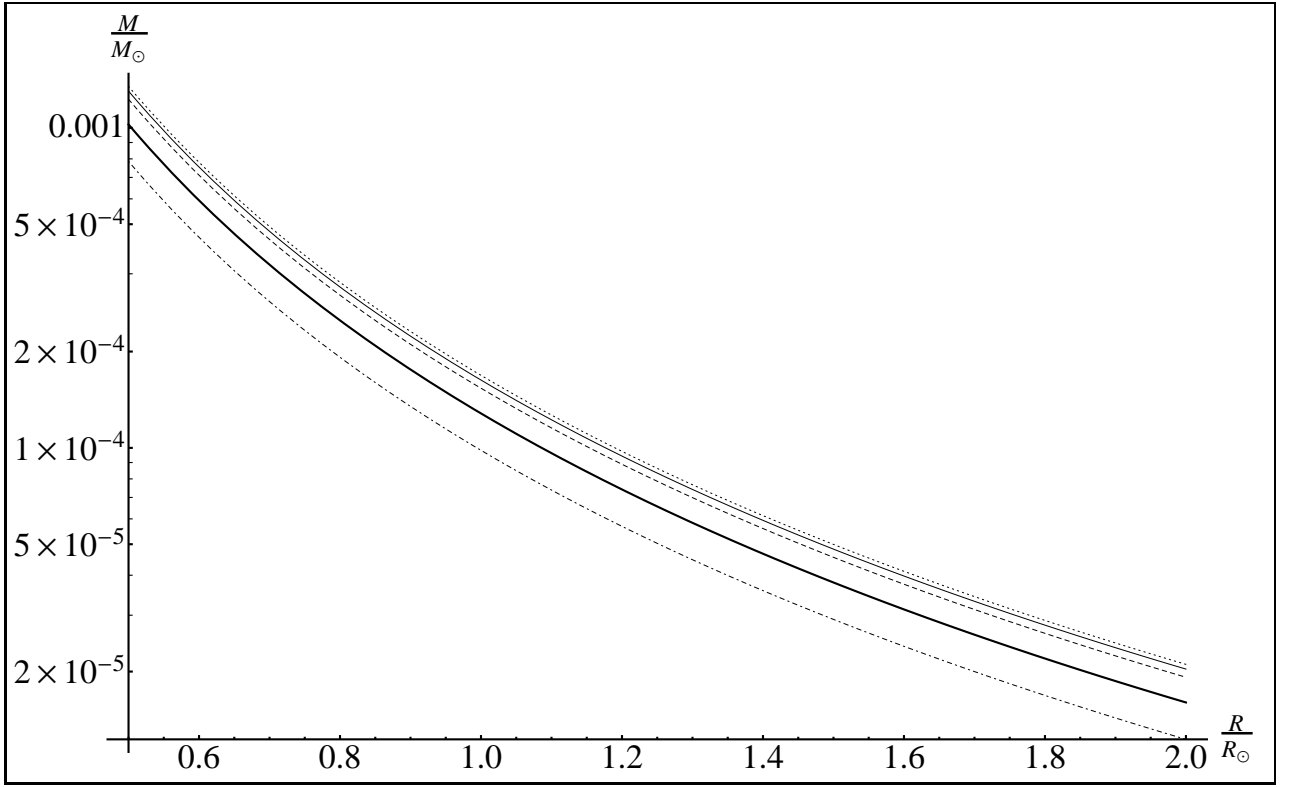} }
\caption{Plot of $\frac{M}{M_\odot}$  as a function of $\frac{R}{R_\odot}$  in the case I, for $n=\frac{3}{2}$  with $N_1=-1.5$ (dotted line), $N_1=0$ (thin line, the isotropic case), $N_1=0.5$ (dashed line), $N_1=1.5$ (thick line) and $N_1=2.5$ (dot--dashed line).}{\label{fig-1}}
\end{figure}

\begin{figure}[tbp]
\centering
\fbox{\includegraphics[scale=0.5]{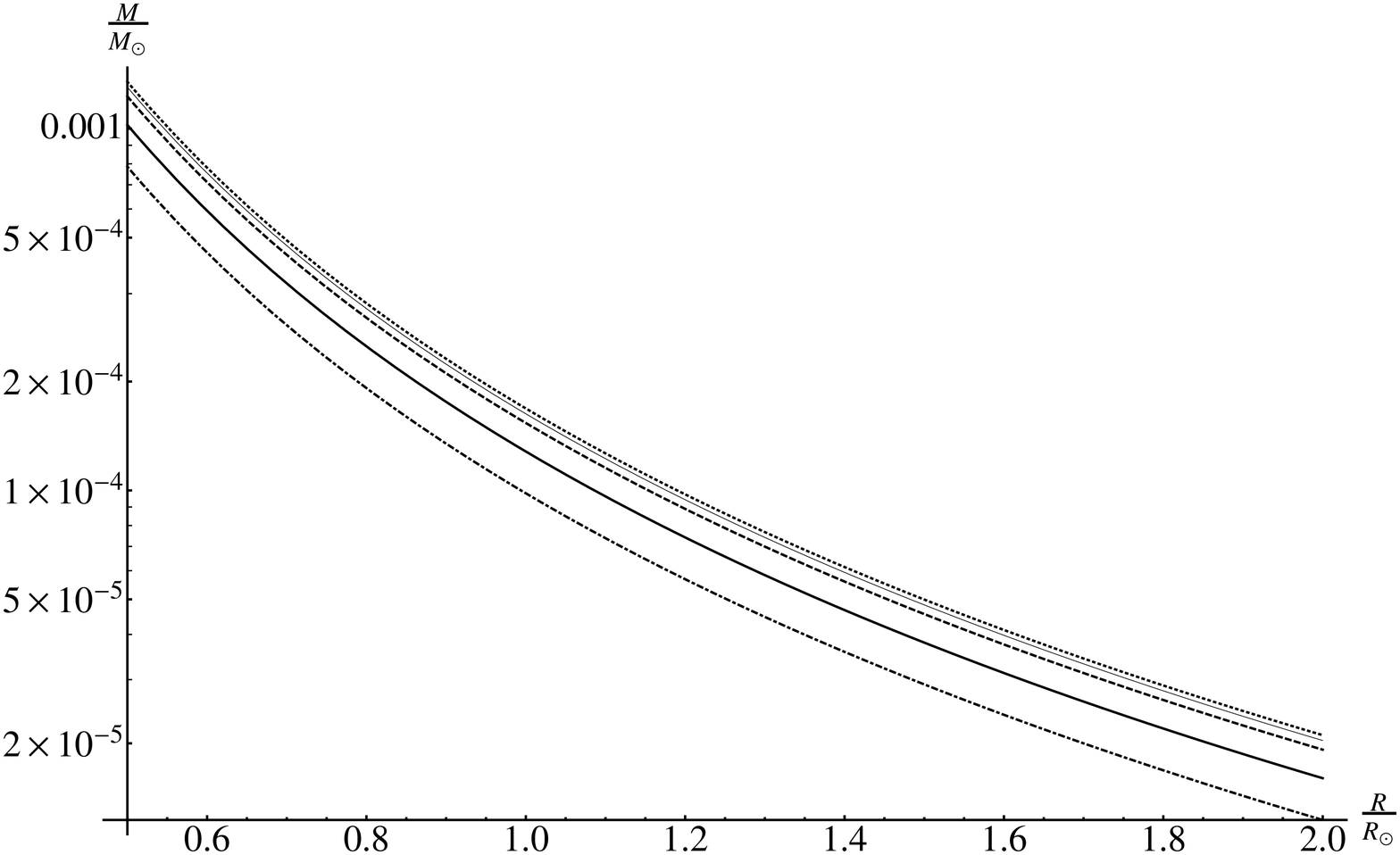} }
\caption{Plot of $\frac{M}{M_\odot}$  as a function of $\frac{R}{R_\odot}$ in the case II, for $n=\frac{3}{2}$  with $N_2=-0.8$ (dotted line), $N_2=0$ (thin line, the isotropic case), $N_2=0.5$ (dashed line), $N_2=1.5$ (thick line) and $N_2=2.5$ (dot--dashed line).}{\label{fig0}}
\end{figure}

Now let us to interprete physically the equation (\ref{A7}). The quantities of physical interest, for definite values of the polytropic index, are the stellar radius and mass. These are:
\begin{equation}
R=\left(\frac{K(n+1)}{4\pi G}\right) ^{\frac{1}{2}}\rho^{\frac{1-n}{2n}}_0\xi_1
\label{rad}\end{equation}
\begin{equation}
M={\int_0}^R 4\pi r^2 \rho dr=4\pi\left(\frac{K(n+1)}{4\pi G}\right)^{\frac{3}{2}}\rho_0 {\int_0}^{\xi_1} \theta^n \xi^2 d\xi
\label{M}\end{equation}
where $\xi_1$ is the first root of $\theta$ function and thus this point defines the surface of the star. Using the anisotropic Lane-Emden equation (\ref{A7}), we get:
\begin{equation}
M=-4\pi\left(\frac{K(n+1)}{4\pi G}\right)^{\frac{3}{2}}\rho_0 {\int_0}^{\xi_1} \frac{d}{d\xi}\left(\xi^2\frac{d\theta}{d\xi}-\frac{2}{P_0(n+1)}\Delta \xi \theta^{-n}\right)  d\xi
\label{MM}\end{equation}
which with the boundary conditions $\theta_0=1$, the relation (\ref{A9}) and the definition of surface $\xi_1$ gives: 
\begin{equation}
M=4\pi\left(\frac{K(n+1)}{4\pi G}\right)^{\frac{3}{2}}\rho_0\left|\xi^2\frac{d\theta}{d\xi}\right|_{\xi_1} 
\label{MMM}\end{equation}
This has the known form of the mass relation for isotropic star. Eliminating the central mass density between (\ref{rad}) and (\ref{MMM}) we have:
\begin{equation}
M R^{\frac{3-n}{n-1}}=4\pi\left(\frac{K(n+1)}{4\pi G}\right)^{\frac{n}{n-1}}{\xi_1}^{\frac{n+1}{n-1}}\left|\frac{d\theta}{d\xi}\right|_{\xi_1} 
\label{MR}\end{equation}
knowing the polytropic index, $\left|\frac{d\theta}{d\xi}\right|_{\xi_1}$ can be obtained from (\ref{A11}) or (\ref{AG}) numerically and thus the constant on the right hand side of equation (\ref{MR}) is found. Therefore the anisotropy factor doesn't appear explicitly in the mass or the mass--radius relation of a star. The effect of anisotropy is included in $\theta$ function through it's equation (parameter $N_1$ in (\ref{A11}) and variable $\eta$ in (\ref{AG})).  Figures (\ref{fig-1}) and (\ref{fig0}) show the mass-radius diagrams for low density white dwarfs with $n=\frac{3}{2}$ obtained from equations (\ref{A11}) and (\ref{AG}) respectively.
\section{Exact analytical solutions of anisotropic Lane-Emden equation}
\subsection{Case I}
\subsubsection{$n=0$}
With $n=0$, the density is constant, $\rho=\rho_0$, and equation (\ref{A11}) takes the following form
\begin{equation}
\theta''+\frac{S-1}{\xi}\theta'+1=0
\label{A13}\end{equation}
in which $S=N_1+N$, where solution is
\begin{equation}
\theta=\left\{
  \begin{array}{ll}
     -\frac{\xi^2}{2S}+1, & \hbox{$S > 1$} \\
\\
  -\frac{\xi^2}{2S}+A\frac{\xi^{2-S}}{2-S}+1  , & \hbox{$S\leq 1, S\neq0$}
  \end{array}
\right.
\label{A14}\end{equation}
in which $A$ is a constant. In the above solution we have also implemented the boundary condition $\theta(0)=1$ and only those terms that lead to the non-singular solutions have been written.

\begin{figure}[tbp]
\centering
\fbox{\includegraphics[scale=0.5]{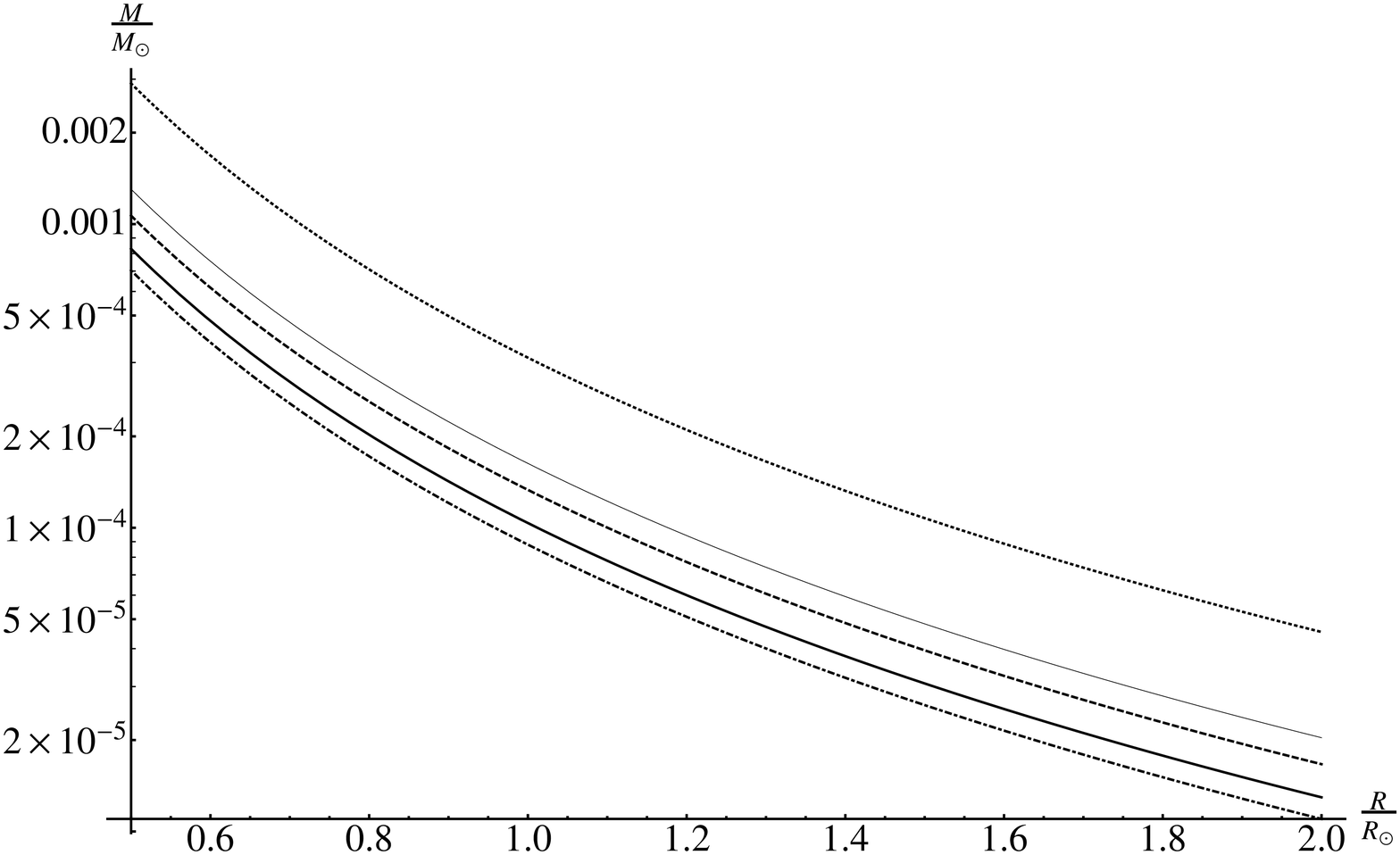} }
\caption{Plot of $\Delta$ for $S>1$ in three dimensions with $S=4$ (thick line), $S=5$ (dotted line), $S=6$ (dashed line), $S=7$ (dot--dashed line).}{\label{fig1}}
\end{figure}

We can now read $\Delta$ from (\ref{A12}) considering the boundary condition (\ref{A9}):
\begin{equation}
\Delta=\left\{
  \begin{array}{ll}
         \frac{(S-N)P_0}{2SN}\xi^2      &  \hbox{$S>1$} \\
 \\   
    -\frac{(S-N)P_0}{2}\left(-\frac{\xi^2}{NS}+\frac{A}{N-S
        }\xi^{2-S}\right)  &  \hbox{$S\leq 1, S\neq 0,  S\neq N$} \\
 \\  0     &  \hbox{$N=S=1$} \\ 
   \end{array}
\right.
\label{AI}\end{equation}
We see that, to satisfy the boundary conditions, we have to have  $\Delta(0)=0$. The resulting  $\Delta$ function is plotted in Fig.(\ref{fig1}). 
 Also the radial and tangential pressures are determined 
 from the expressions (\ref{A6}),  (\ref{A14}) and (\ref{AI}).

\subsubsection{$n=1$}
\begin{figure}[tbp]
\centering
\fbox{\includegraphics[scale=0.5]{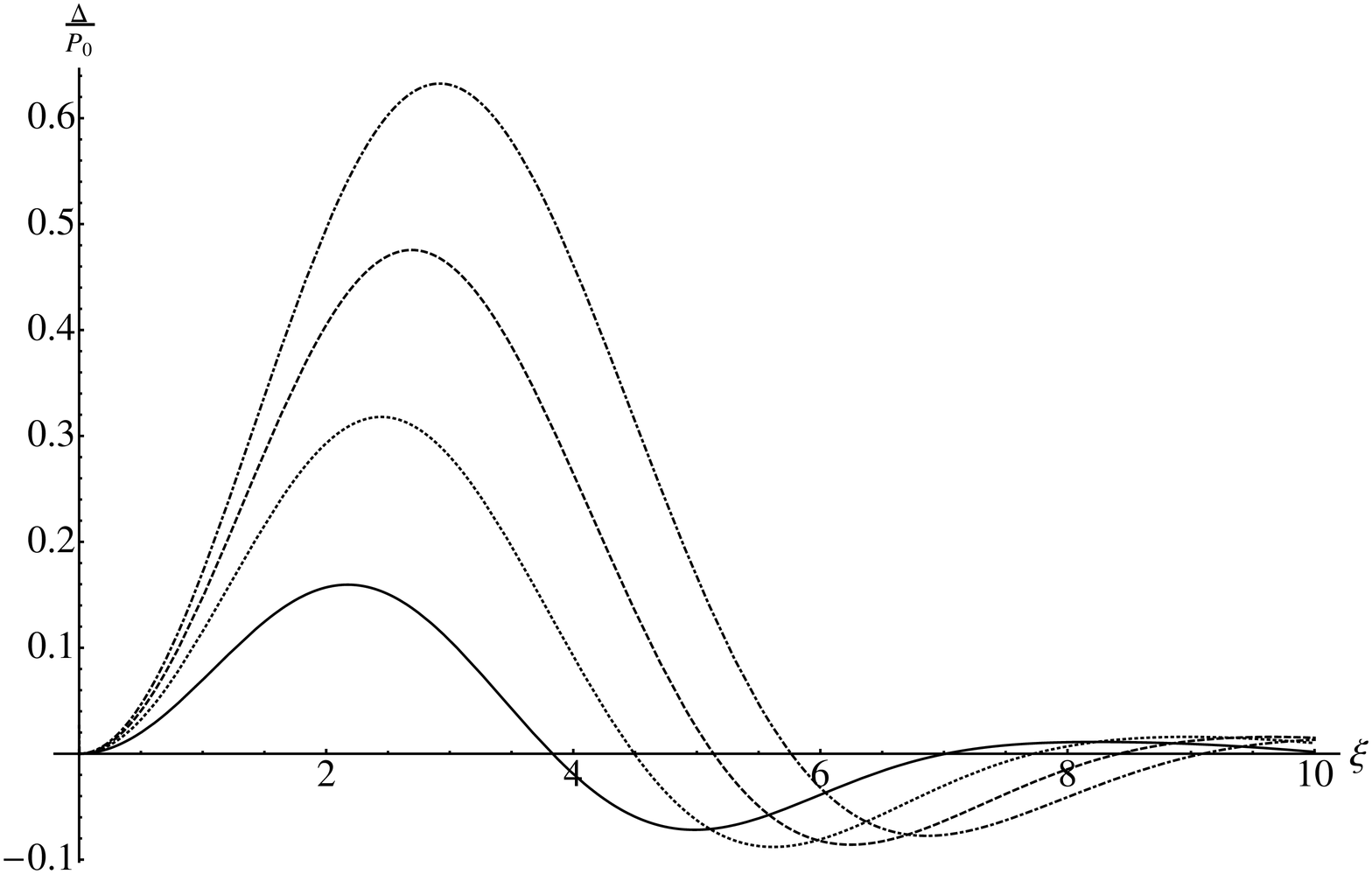} }
\caption{Plot of $\Delta$ for $n=1$ in three dimensions  with $S=4$ (thick line), $S=5$ (dotted line), $S=6$ (dashed line), $S=7$ (dot--dashed line).}{\label{fig2}}
\end{figure}

In this case, the differential equation (\ref{A11}) reduces to 
\begin{equation}
\theta''+\frac{(S-1)}{\xi^2} \theta' + \theta=0
\label{A27}\end{equation}
The solutions of (\ref{A27}) are given in term of Bessel functions as follows
\begin{equation}
\theta(\xi)=\xi^{-\frac{S-2}{2}}\left[C_1 J_{\frac{S-2}{2}}(\xi)+C_2Y_{\frac{S-2}{2}}(\xi)\right]
\label{A28}\end{equation}
where $C_1$ and $C_2$ are integration constants. Imposing the condition $\theta(0)=1$, $C_2$ vanishes and we get
\begin{equation}
 \theta(\xi)=\Gamma \left(\frac{S}{2}\right)\left(\frac{\xi}{2}\right)^{-\frac{S-2}{2}}J_{\frac{S-2}{2}}(\xi),\hspace{1cm}S>2
\label{A29}\end{equation}
in which the constant $C_1$ is expressed in terms of Gamma function, $\Gamma\left(\frac{S}{2}\right)$. Substituting (\ref{A29}) into (\ref{A12}) leads to
\begin{equation}
\label{A31}
\Delta(\xi)=\frac{2^{\frac{S}{2}}}{8}(S-N) P_{0}\xi^{3-\frac{s}{2}}J_{\frac{S-2}{2}}(\xi)\Gamma^2(\frac{S}{2})\Gamma(\frac{N}{2})F\left(\frac{N}{2};1+\frac{S}{2},1+\frac{N}{2};-\frac{\xi^2}{4}\right)
\end{equation}
where $F(a,b,c,x)$ is the hypergeometric function. These results also satisfy the boundary condition (\ref{A9}) with $\theta'(0)=\Delta(0)=0$. The function (\ref{A31}) is depicted in the Fig. (\ref{fig2}).

\subsubsection{$n=\frac{S+2}{S-2}$}
\begin{figure}[tbp]
\centering
\fbox{\includegraphics[scale=0.5]{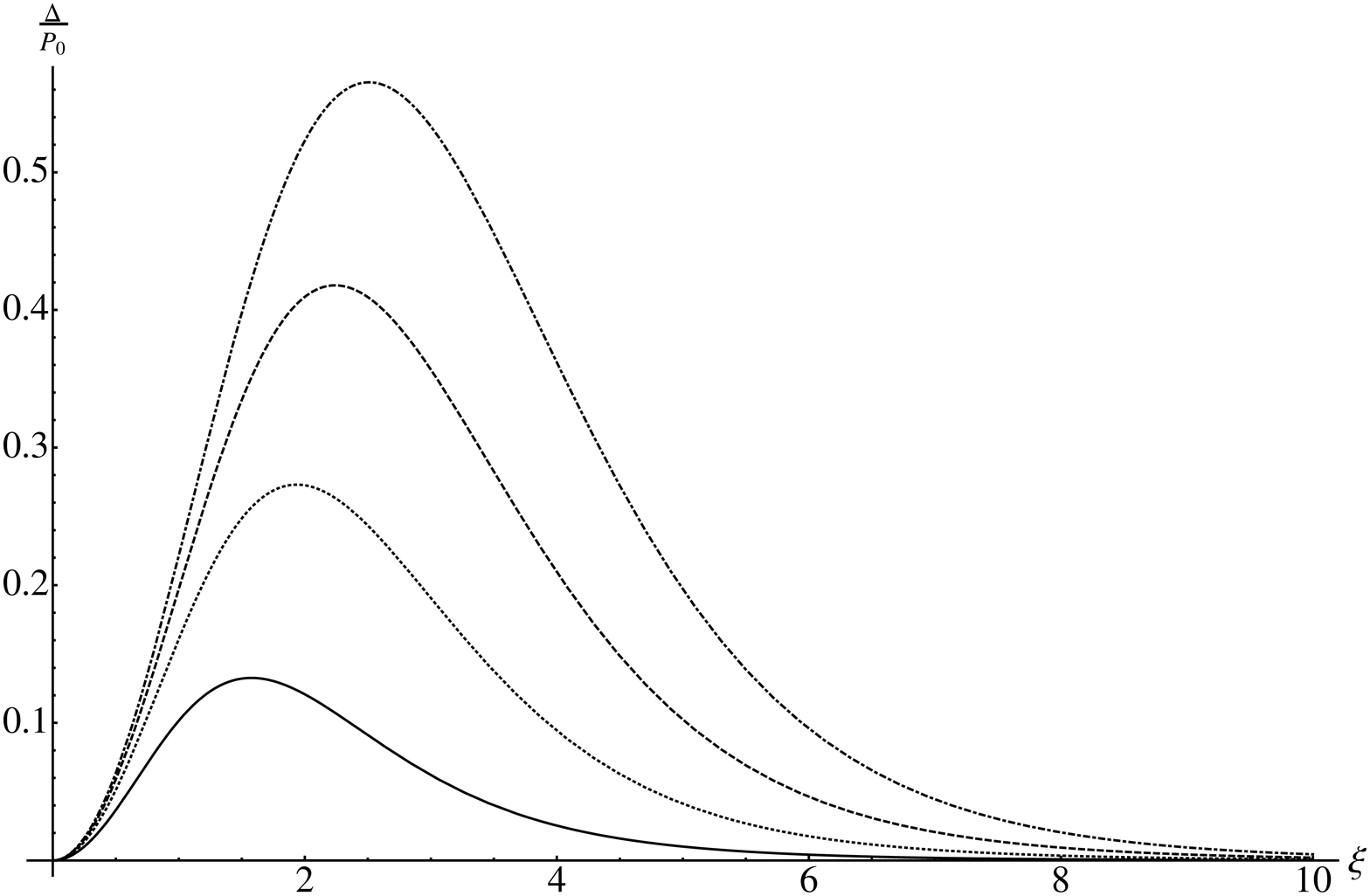} }
\caption{Plot of $\Delta$ for $n=\frac{S+2}{S-2}$ in three dimensions with $S=4$ (thick line), $S=5$ (dotted line), $S=6$ (dashed line), $S=7$ (dot--dashed line).}{\label{fig3}}
\end{figure}

In this case, if one takes
\begin{equation}
\theta(\xi)=\left(\frac{4}{(n-1)^2}\right)^{n-1}\xi^{\frac{2}{1-n}}z(\xi),\hspace{1cm}\xi =e^{-t}
\label{A35}\end{equation}
then equation (\ref{A11}) reduces to
\begin{equation}
\frac{d^2z}{dt^2}=4\frac{z\mp z^n}{(n-1)^2}
\label{A37}\end{equation}
The above equation can be integrated as follows:
\begin{equation}
\frac{1}{2}\left(\frac{dz}{dt}\right)^2=\frac{4}{(n-1)^2}\left[\frac{z^2}{2}\mp \frac{z^{n+1}}{n+1}\right]+C
\label{A38}\end{equation}
Considering $n>1$ ($S>2$), turning back to the original variables and using equations (\ref{A35}) and (\ref{A38}), a simple calculation shows that the initial conditions on $\theta$ leads to $z = 0$ and $dz/dt=0$ at $\xi=0$.  This gives $C=0$. Writing the above equation in terms of the original variables, we have:
\begin{equation}
\frac{2\theta \theta'}{n-1}+\xi \frac{\theta'^2}{2}+\xi \frac{\theta^{n+1}}{n+1}=0
\label{A39}\end{equation}
We proceed with integrating the above equation
\begin{equation}
\xi \frac{\theta^{\frac{n+1}{2}}}{\theta'}=D
\label{A41}\end{equation}
The constant $D$ might be determined taking into account (\ref{A9}) and $\theta(0)=1$
\begin{equation}
\frac{1}{D}=\lim_{\xi \to 0}\theta^{-\frac{S}{S-2}}\frac{\theta'}{\xi}=-\frac{1}{N}+\frac{2}{P_0(n+1)}\lim_{\xi \to 0}\frac{\Delta(\xi)}{\xi^2}
\label{A42}\end{equation}
As we see, the constant $D$ depends on the $\Delta$ function near the origin. The differential equation (\ref{A41}) has the following solution
\begin{equation}
\theta=\left[1-\frac{\xi^2}{D(S-2)}\right]^{-\frac{S-2}{2}}
\label{A43}
\end{equation}
Employing (\ref{A12}), we get the following expression for the difference between the radial and tangential pressures
\begin{equation}
\Delta(\xi)=-\frac{S P_0 (S-N)}{ND(S-2)}\xi^2\left[1-\frac{\xi^2}{D(S-2)}\right]^{-\frac{S+2}{2}} F\left(\frac{N}{2},\frac{S}{2},1+\frac{N}{2},\frac{\xi^2}{D(S-2)}\right)
\label{A44}
\end{equation}
Now, we are in the position to fix the constant $D$. Combining (\ref{A42}) and (\ref{A44}), one gets $D=-S$. 
The resulting $\Delta$ is shown in Fig. (\ref{fig3}).
\subsection{Case II}
\subsubsection{$n=0$}
\begin{figure}[tbp]
\centering
\fbox{\includegraphics[scale=0.5]{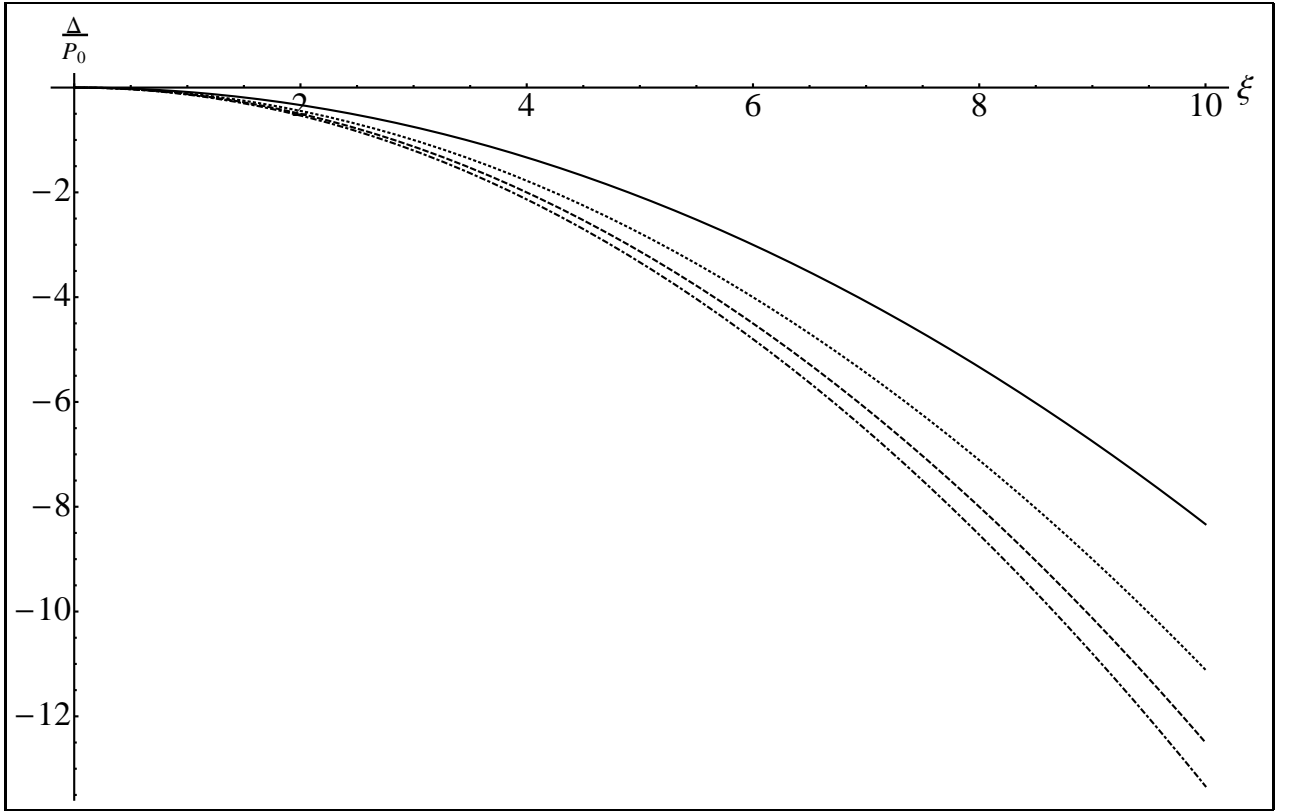} }
\caption{Plot of $\Delta$ for $n=0$ in three dimensions with $N_2=1$ (thick line), $N_2=2$ (dotted line), $N_2=3$ (dashed line), $N_2=4$ (dot--dashed line).}{\label{fig4}}
\end{figure}

The solutions of differential equations (\ref{AG}) and (\ref{AH}) satisfying the boundary conditions are:
\begin{equation}
\theta=\left\{
  \begin{array}{ll}
     -\frac{N_2+1}{2N}\xi^2+A\sqrt{N_2+1}\xi+1, & \hbox{$N=1$} \\
\\
  -\frac{N_2+1}{2N}\xi^2+1  , & \hbox{$N > 1$}
  \end{array}
\right.
\label{AJJ}
\end{equation}
\begin{equation}
\Delta=\left\{
  \begin{array}{ll}
     -\frac{P_0 N_2}{2(N_2+1)N}\xi^2, & \hbox{$N\geq 2$} \\
\\
  -\frac{P_0N_2}{2(N_2+1)}(\xi^2-\frac{\sqrt{(N_2+1)^3}}{N_2}A\xi)  , & \hbox{$N =1$}
  \end{array}
\right.
\label{AJ}
\end{equation}
The behaviour of $\Delta$ is shown in Fig. (\ref{fig4}).
\subsubsection{$n=1$}
\begin{figure}[tbp]
\centering
\fbox{\includegraphics[scale=0.5]{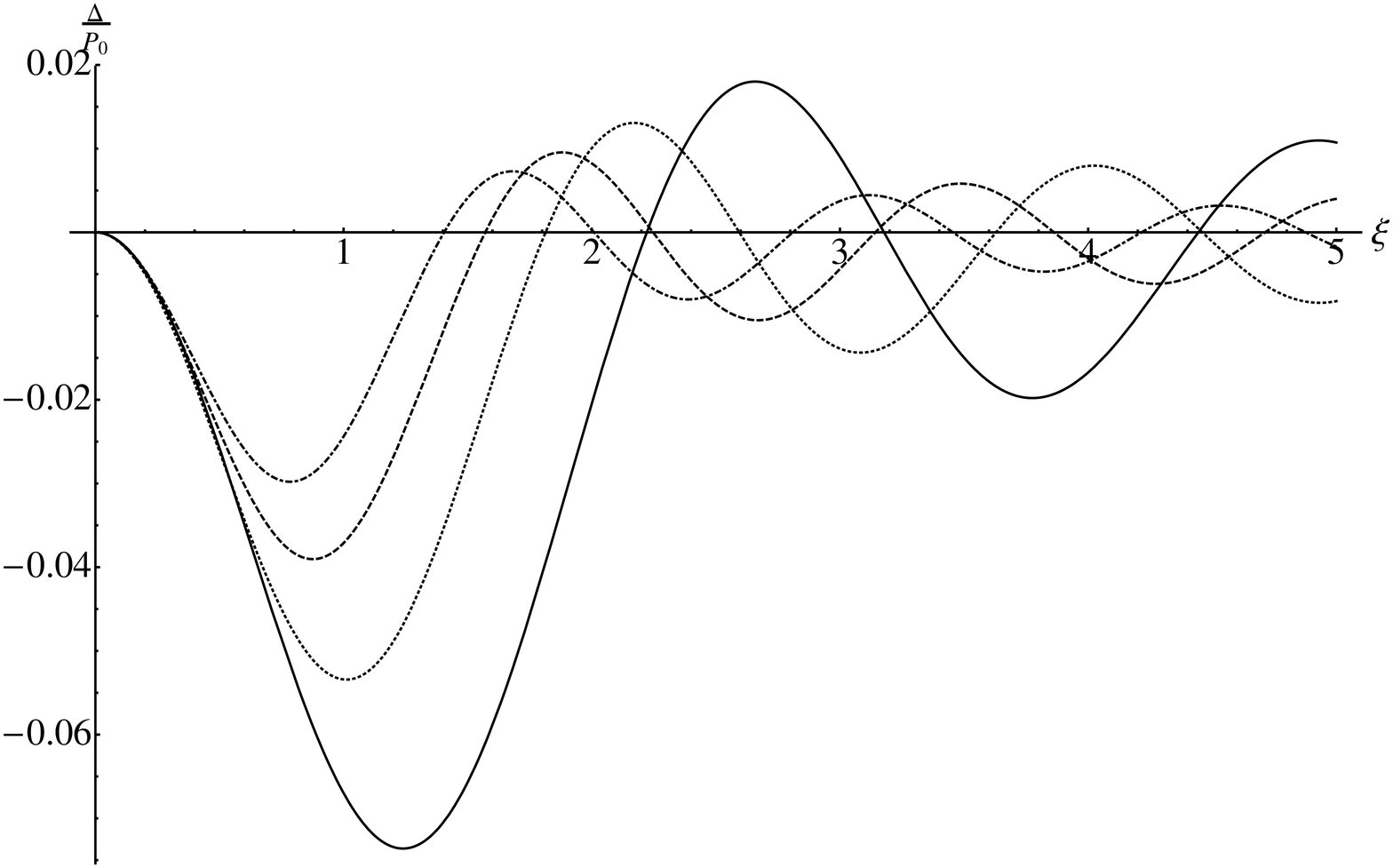} }
\caption{Plot of $\Delta$ for $n=1$ in three dimensions with $N_2=1$ (thick line), $N_2=2$ (dotted line), $N_2=3$ (dashed line), $N_2=4$ (dot--dashed line).}{\label{fig5}}
\end{figure}

The situation for this case is also similar to the previous one and the substitution of $S$ and $\xi$ by $N$ and $\eta$ in (\ref{A29}) produces the solution of (\ref{AG}) with $n=1$. This outcome with the integration (\ref{AH}) give the difference between the radial and tangential pressures: 
\begin{equation}
\Delta(\xi)=-2^{N-2}\frac{P_0 N_2}{(N_2+1)^{\frac{1}{2}(N+1)}}\Gamma^2(\frac{N}{2})\xi^{3-N}J_{\frac{N-2}{2}}(\sqrt{N_2+1}\xi)J_{\frac{N}{2}}(\sqrt{N_2+1}\xi)
\label{AP}
\end{equation}
which are depicted in Fig. (\ref{fig5}).
\subsubsection{$n=\frac{N+2}{N-2}$}
\begin{figure}[tbp]
\centering
\fbox{\includegraphics[scale=0.5]{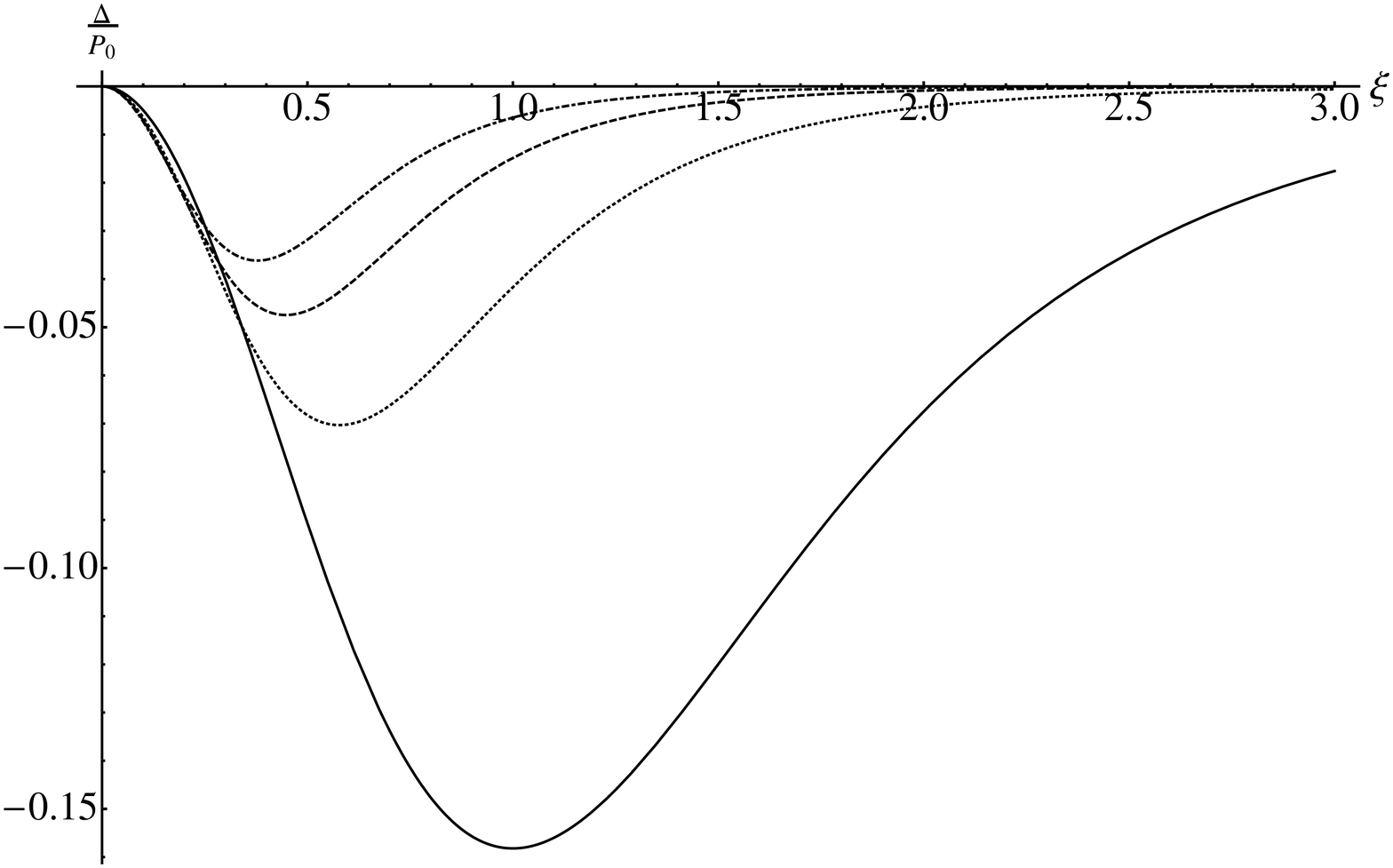} }
\caption{Plot of $\Delta$ for $n=\frac{N+2}{N-2}$ in three dimensions with $N_2=1$ (thick line), $N_2=2$ (dotted line), $N_2=3$ (dashed line), $N_2=4$ (dot--dashed line).}{\label{fig6}}
\end{figure}

Similar to the two former cases, in this situation the solution to the (\ref{AG}) with $n=\frac{N+2}{N-2}$ might be obtained by changing $S$ and $\xi$ to $N$ and $\eta$ in (\ref{A43}). 
\begin{equation}
\theta=\left[1-\frac{N_2+1}{D(N-2)}\xi^2\right]^{-\frac{N-2}{2}} \hspace{1cm}N> 2
\label{AR}
\end{equation}
Putting the above result in (\ref{AH}) gives:
\begin{equation}
\Delta=-\frac{NP_0N_2}{(N_2+1)(N-2)N}\left[1-\frac{1+N_2}{D(N-2)}\xi^2\right]^{-(N+1)}\xi^2.
\label{AS}\end{equation}
Combining (\ref{A42}) and (\ref{AS}), one gets $D=-\frac{N(N_2+1)}{2N_2-1}$. The graph of (\ref{AS}) is is plotted in Fig. (\ref{fig6}) for various values of $N_2$.
\section{Integral Theorems}
Chandrasekhar in his famous book \cite{chandra} has discussed some inequalities for the physical quantities describing a star in the Newtonian gravitational equilibrium. In this section we will extend some of his results for anisotropic stars. The results are general and are not restricted to the polytropic case. 

Let $m(r)$ be the mass contained inside radius $r$, then 
\begin{equation}
m'=4\pi\rho r^2
\label{c1}
\end{equation}
Equation (\ref{A1}) can now be written as
\begin{equation}
P_r'=-\frac{Gm\rho}{r^2}+\frac{2\Delta}{r}
\label{c2}
\end{equation} 
where we have used the definition of gravitational potential.
From (\ref{c1}) and (\ref{c2}) one can get the following equation:
\begin{equation}
\frac{1}{r^2}\left[\frac{r^2}{\rho}\left(P_r'-\frac{2\Delta}{r}\right)\right]' =-4\pi G\rho
\end{equation}
These equations can be used to prove the following theorems:

\underline{\textbf{Theorem 1:}}

For any equilibrium configuration the function 
\begin{equation}
I=P_r+ \frac{Gm^2}{8\pi r^4}-2\int_0^rd\tilde{r}\frac{\Delta}{\tilde{r}}
\end{equation}
does not increases outward.

\textbf{Proof:}
Let's calculate
\begin{equation}
I'=P_r'+\frac{Gmm'}{4\pi r^4}-\frac{Gm^2}{2\pi r^5}-\frac{2\Delta}{r}
\end{equation}
Then by (\ref{c2})
\begin{equation}
I'=-\frac{Gm^2}{2\pi r^5}\leq0
\end{equation}
QED.

\hskip1cm\underline{\textbf{Corollary :}}

For the central pressure, we have
\begin{equation}
P_0>P_r+ \frac{Gm^2}{8\pi r^4}-2\int_0^rd\tilde{r}\frac{\Delta}{\tilde{r}}>\frac{GM^2}{8\pi R^4}-2\int_0^Rd\tilde{r}\frac{\Delta}{\tilde{r}}
\label{zarb}
\end{equation}
in which $M=m(R)$ is the mass of the star. The last term in the right hand side shows a lower bound on $P_0$. Moreover for any arbitrary radius, the above relation leads to:
\begin{equation}
P_r>\frac{G}{8\pi}\left( \frac{M^2}{R^4}- \frac{m^2}{r^4}\right)-2\int_r^Rd\tilde{r}\frac{\Delta}{\tilde{r}}
\end{equation}

\underline{\textbf{Theorem 2:}}

For any equilibrium configuration

\begin{equation}
I_\nu\equiv \int_0^R dm \frac{Gm}{\tilde{r}^\nu}=4\pi\int_0^R d\tilde{r}\tilde{r}^{3-\nu}\left[
\left(4-\nu\right)P_r+2\Delta\right]
\label{ho}
\end{equation}
if $\nu<4$.

\textbf{Proof:}

From (\ref{c2}) we have
\begin{equation}
\frac{Gmm'}{4\pi r^4}=\frac{2\Delta}{r}-P_r'
\end{equation}
Multiplying this by $r^{4-\nu}$ and then integrating from $r$ to $R$ gives:
\begin{equation}
I_\nu=8\pi\int_0^R d\tilde{r}\tilde{r}^{\nu-3}\Delta-4\pi\int_0^R dP_r\tilde{r}^{4-\nu}
\label{j}
\end{equation}

Integrating by parts the first integral, leads to (\ref{ho}). QED.

For $\nu=4$, equation (\ref{j}) gives:
\begin{equation}
I_4=8\pi\int_0^Rd\tilde{r}\frac{\Delta}{\tilde{r}}+4\pi P_0
\end{equation}
which also has a lower bound equal to $\frac{GM^2}{2R^4}$ according to (\ref{zarb}).

The gravitational potential energy $\Omega$ of the configuration is $-I_1$. According to (\ref{j}) it is given by:
\begin{equation}
\Omega=-\int_0^R dV \left(3P_r+2\Delta \right)
\end{equation}
Denoting the mean value of gravitational acceleration by $\bar{g}$, we have:
\begin{equation}
M\bar{g}=\int_0^R dm\frac{Gm}{r^2}=I_2
\end{equation}
and hence by (\ref{j}):
\begin{equation}
M\bar{g}=8\pi\int_0^R d\tilde{r}\tilde{r}\left(P_r+\Delta \right)=8\pi\int_0^R d\tilde{r}\tilde{r}P_t
\end{equation}
This shows that for an anisotropic star only the tangential pressure contributes in finding the mean value of acceleration.

\underline{\textbf{Theorem 3:}}
For any equilibrium configuration
\[
\pi\nu P_0R^{4-\nu}+\frac{(4-\nu) GM^2}{8 R^\nu}+2\pi\nu(4-\nu)\int_0^R d\tilde{r}\tilde{r}^{3-\nu}\int_0^R d\tilde{r}\frac{\Delta}{\tilde{r}}
\]
\begin{equation}
>2\pi\nu\int_0^R d\tilde{r}\frac{\Delta}{\tilde{r}^{\nu-3}}+I_\nu>\frac{GM^2}{2R^\nu}
\label{*}
\end{equation}

\textbf{Proof:}

By theorem 1:
\begin{equation}
\frac{G M^2}{8\pi R^4}-\frac{G m^2}{8\pi r^4}-2\int_0^Rd\tilde{r}\frac{\Delta}{\tilde{r}}+2\int_0^rd\tilde{r}\frac{\Delta}{\tilde{r}}<P_r<P_0-\frac{Gm^2}{8\pi r^4}+2\int_0^rd\tilde{r}\frac{\Delta}{\tilde{r}}
\end{equation}
and by theorem 2:
\[
4\pi(4-\nu)\int_0^Rd\tilde{r}\tilde{r}^{3-\nu}\left[\frac{G M^2}{8\pi R^4}-\frac{G m^2}{8\pi r^4}-2\int_0^Rd\tilde{r}\frac{\Delta}{\tilde{r}}+2\int_0^rd\tilde{r}\frac{\Delta}{\tilde{r}}\right]
\]
\begin{equation}
<I_\nu +8\pi\int_0^R d\tilde{r}\tilde{r}^{3-\nu}\Delta<4\pi(4-\nu)\int_0^Rd\tilde{r}\tilde{r}^{3-\nu}\left[P_0-\frac{Gm^2}{8\pi r^4}+2\int_0^rd\tilde{r}\frac{\Delta}{\tilde{r}}\right]
\end{equation}
This inequality can be written as:
\[
4\pi P_0 R^{4-\nu}+8\pi\nu(4-\nu)\int_0^R d\tilde{r}\tilde{r}^{3-\nu}\int_0^r d\tilde{r}\frac{\Delta}{\tilde{r}}
\]
\begin{equation}
>I_\nu +8\pi\int_0^R d\tilde{r}\tilde{r}^{3-\nu}\Delta+\frac{(4-\nu)}{2}\int_0^R d\tilde{r} \frac{Gm^2}{\tilde{r}^{\nu+1}}>\frac{GM^2}{2R^\nu}
\label{y}
\end{equation}
Inserting 
\begin{equation}
\int_0^R d\tilde{r} \frac{Gm^2}{\tilde{r}^{\nu+1}}=\frac{GM^2}{R^\nu}-2I_\nu
\end{equation}
in (\ref{y}) and simplifying it, we get (\ref{*}). QED.

\hskip1cm\underline{\textbf{Corollary :}}
Setting $\nu=1$, we have
\begin{equation}
\left(\pi P_0 R^3+\frac{3GM^2}{8R}+6\pi\int_0^R d\tilde{r}\tilde{r}^2\int_0^r d\tilde{r}\frac{\Delta}{\tilde{r}}\right) >2\pi\int_0^R d\tilde{r}\tilde{r}^2\Delta-\Omega>\frac{GM^2}{2R}
\end{equation}
which shows the upper and lower bounds of potential energy, $-\Omega$.
\section{Concluding remarks}
In this paper we discussed how the anisotropy factor modifies the Lane-Emden equation and homology theorem. We obtained some theorems governing the characteristic functions of anisotropic star. These are the extension of Chandrasekhar's theorems. We performed a procedure to find the anisotropy factor, the radial pressure and the density functions exactly satisfying the Lane-Emden equation. We had two cases. In the first case the effect of anisotropy is to change the dimmension of space from $N$ to $N+N_1$ (which can be a non--integer number). For the second case, anisotropy shows itself as a rescaling of the radial coordinate in the function $\theta$.

\begin{figure}[tbp]
\centering
\fbox{\includegraphics[scale=0.5]{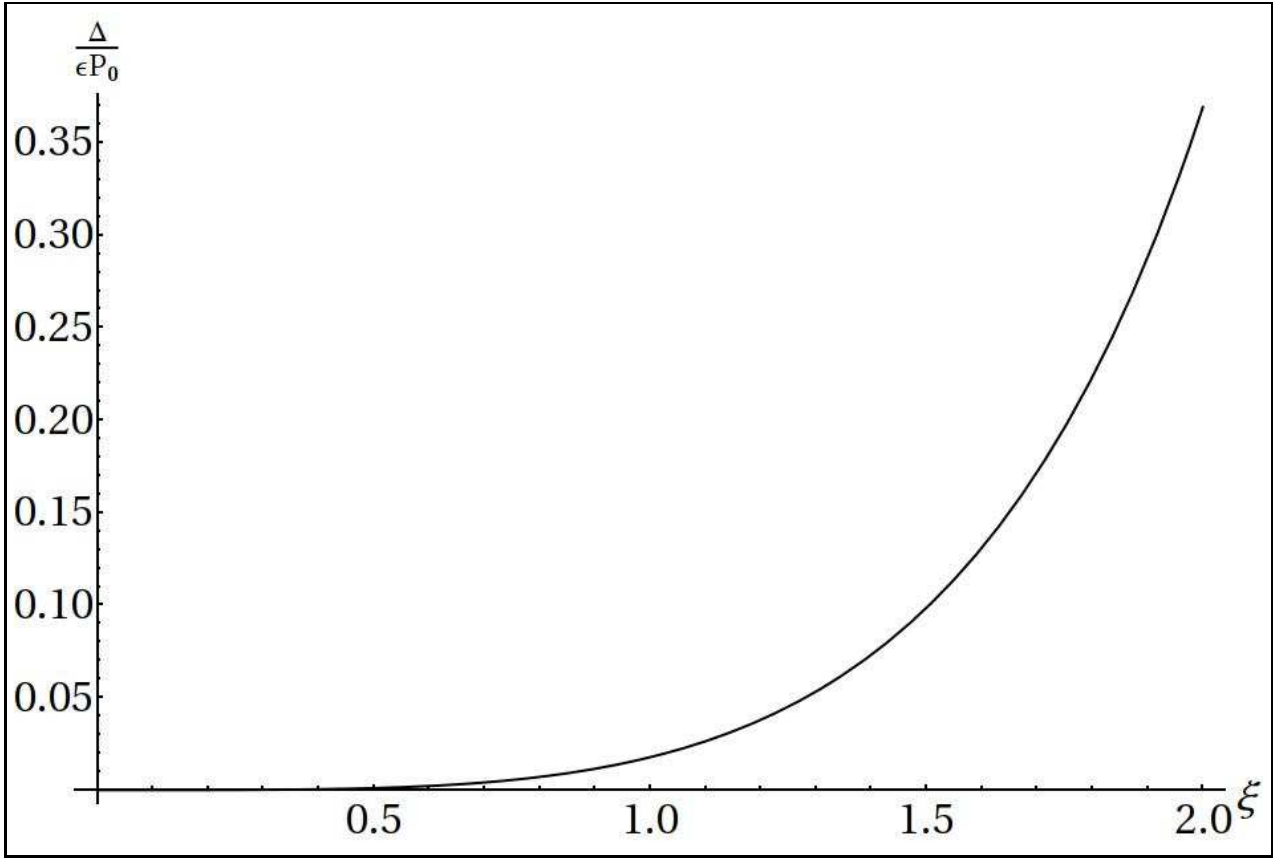}}
\caption{Plot of $\Delta$ for the analytical solution close to the exact solution ($n_0=0$, $N=3$) as explained in the text.}{\label{fig7}}
\end{figure}

Moreover it is straightforward to find some approximate analytical solutions of the polytropic stars with anisotropic pressure. For example, let's assume that the including anisotropy factor is equivalent to a slight modification of the  polytropic index from it's value in the absence of anisotropy, $n_0$. This means that if an exact solution  $\theta_0$ of isotropic Lane-Emden equation is known:
\begin{equation}
\theta_0''+\frac{N-1}{\xi}\theta'_0=-\theta_0^{n_0}.
\label{unper}
\end{equation}
Inclusion of anisotropy factor leads to:
\begin{equation}
\theta''+\frac{N-1}{\xi}\theta'-\frac{2}{P_0 (n_0+1)\xi\theta^{n_0}}\left[\Delta'+\frac{1}{\xi}\Delta-n_0\frac{\theta'}{\theta}\Delta\right]=-\theta^{n_0}
\label{badelta}
\end{equation}
which is equivalent to perturbing $\theta_0$ and  $n_0$ in equation (\ref{unper}) in the way that $\theta=\theta_0+\epsilon\theta_*$ and $n=n_0+\epsilon$ satisfy (\ref{unper}):
\begin{equation}
\theta''+\frac{N-1}{\xi}\theta'=-\theta^n.
\label{unpert}
\end{equation}
where $\vert\epsilon\vert\ll 1$. Thus we have the following equation for $\Delta$:
\begin{equation}
\frac{2}{P_0 (n_0+1)\xi\theta^{n_0}}\left[\Delta'+\frac{1}{\xi}\Delta-n_0\frac{\theta'}{\theta}\Delta\right]=-\theta^{n}+\theta^{n_0}
\label{deltaeq}
\end{equation}
Knowing $n_0$, the functions $\theta_0$ and thus $\theta$ can be determined from (\ref{unper}) and (\ref{unpert}) and thus $\Delta$ is obtained from (\ref{deltaeq}). With the initial conditions $\theta_0(0) = 1$ and  $\theta_0'(0)$,  $\theta_*(0)$,  $\theta_*'(0) = 0$ and for the case $n_ 0 = 0$ and $N=3$, we get $\theta_ 0 = 1 - \xi^2 /6$. Inserting $\theta_ 0$ into (\ref{unpert}), we find that \cite{sei}:
\begin{equation}
\theta_* = (3 - \frac{\xi^2}{6}) \ln{(1 - \frac{\xi^2}{6})} + \frac{2\sqrt {6}}{\xi} \ln{\frac{\sqrt{6} + \xi}{\sqrt{6} - \xi}} + \frac{5}{18}\xi^ 2 - 4
\end{equation}
where $\xi \leq \sqrt{6}$. Thus:
\begin{equation}
\Delta= -\frac{P_0 \epsilon}{2}\left[\frac{\xi^2}{9} [3\ln{(1 - \frac{\xi^2}{6})-2}] + \frac{4\sqrt {6}}{\xi} \arctan {\frac{\xi}{\sqrt{6}}} - 4\right] 
\end{equation} 
The plot of $\Delta$ is shown in Fig. (\ref{fig7}). 

{\bf Acknowledgments:}
This work was supported by a grant from University of Tehran.


\begin{thebibliography}{99}
\bibitem{her2013}
Herrera L. and Barreto W., \textit{Phys. Rev. D}, \textbf{87}, 087303, 2013
\bibitem{dev}
Dev K. and Gleiser M., \textit{Gen. Relativ. Grav.}, \textbf{35}, 8, 2003
\bibitem{chandra}
S. Chandrasekhar, An Introduction to the study of Stellar Structure, The University of Chicago Press, 1939
\bibitem{horedt}
Horedt G.P., Polytropes: Applications in Astrophysics and Related Fields, Kluwer Academic Publishers, 2004
\bibitem{rod}
Rodrigues H., \textit{Eur. J. Phys.}, \textbf{34}, 667, 2013
\bibitem{her1997}
Herrera L. and Santos N. O., \textit{Phys. Rep.}, \textbf{286}, 53, 1997
\bibitem{velocity}
Binney J. and Tremaine S., Galactic Dynamics, Princeton University Press, Princeton, 2008;\\
Cuddeford P., \textit{Mon. Not. R. Astron. Sot.} \textbf{253}, 414, 1991
\bibitem{fer}
Ralston J. and Smith L., \textit{Astrophys. J.}, \textbf{367}, 54, 1991
\bibitem{non}
Alencar P. and Letelier P., \textit{Phys. Rev. D}, \textbf{34}, 343, 1986
\bibitem{low}
Liu H., Zhang X. and Wen D., \textit{Phys. Rev. D}, \textbf{89}, 104043, 2014
\bibitem{cos}
Cosensa, M., Herrera L., Esculpi M. and Witten L., \textit{Phys. Rev. D}, \textbf{25}, 2527, 1982;\\
Cosensa, M., Herrera L., Esculpi M. and Witten L., \textit{J. Math. Phys.}, \textbf{22}, 118, 1981
\bibitem{sei}
Seidov, Z. F. and Kuzakhmedov, R. Kh., \textit{Astron. Zh.}, \textbf{55}, 1250, 1978
\end{thebibliography}
\end{document}